\documentclass{ws-ijmpa}
\usepackage[super,compress]{cite}
\usepackage{color}
\usepackage{graphicx}
\def\Journal#1#2#3#4{{#1} {\bf #2}, #3 (#4)}

\def\PLB{{\em Phys. Lett.}  B}

\begin{document}
\markboth{Authors' Names}{$\tau$ mass and $R$ value measurements at BES}

%
\catchline{}{}{}{}{}
%

\title{$\tau$ mass and $R$ value measurements at BES
}

\author{Zhipeng ZHENG
}

\address{Institute of High Energy Physics, Chinese Academy of Sciences\\
Beijing, 100049, China\\
zhengzp@ihep.ac.cn}

\author{Shuangshi FANG}

\address{Institute of High Energy Physics, Chinese Academy of Sciences\\
Beijing, 100049, China\\
fangss@ihep.ac.cn}

\author{Guangshun HUANG}

\address{University of Science and Technology of China\\
Heifei, 230026, China\\
hgs@ustc.edu.cn
}

\maketitle

\begin{history}
\received{Day Month Year}
\revised{Day Month Year}
\end{history}

\begin{abstract}
A comprehensive review of the measurements of $\tau$ lepton
mass and $R$ values in the energy region between $2-5$ GeV, achieved at the
BES experiment, is presented.  In addition to the evaluation of their impact
on the test of Standard Model, we also highlighted the present status and
the most recent developments.  In particular, we made an extensive
discussion on the prospects for future improvements at the BESIII
experiment.

\keywords{the BES detector; $\tau$ lepton mass; $R$ value.}
\end{abstract}

\ccode{PACS numbers: 13.35.Dx 14.60.Fg 13.66.Jn}


\section{Introduction}

\begin{figure}[htpb]
\centerline{\includegraphics[width=13.8cm]{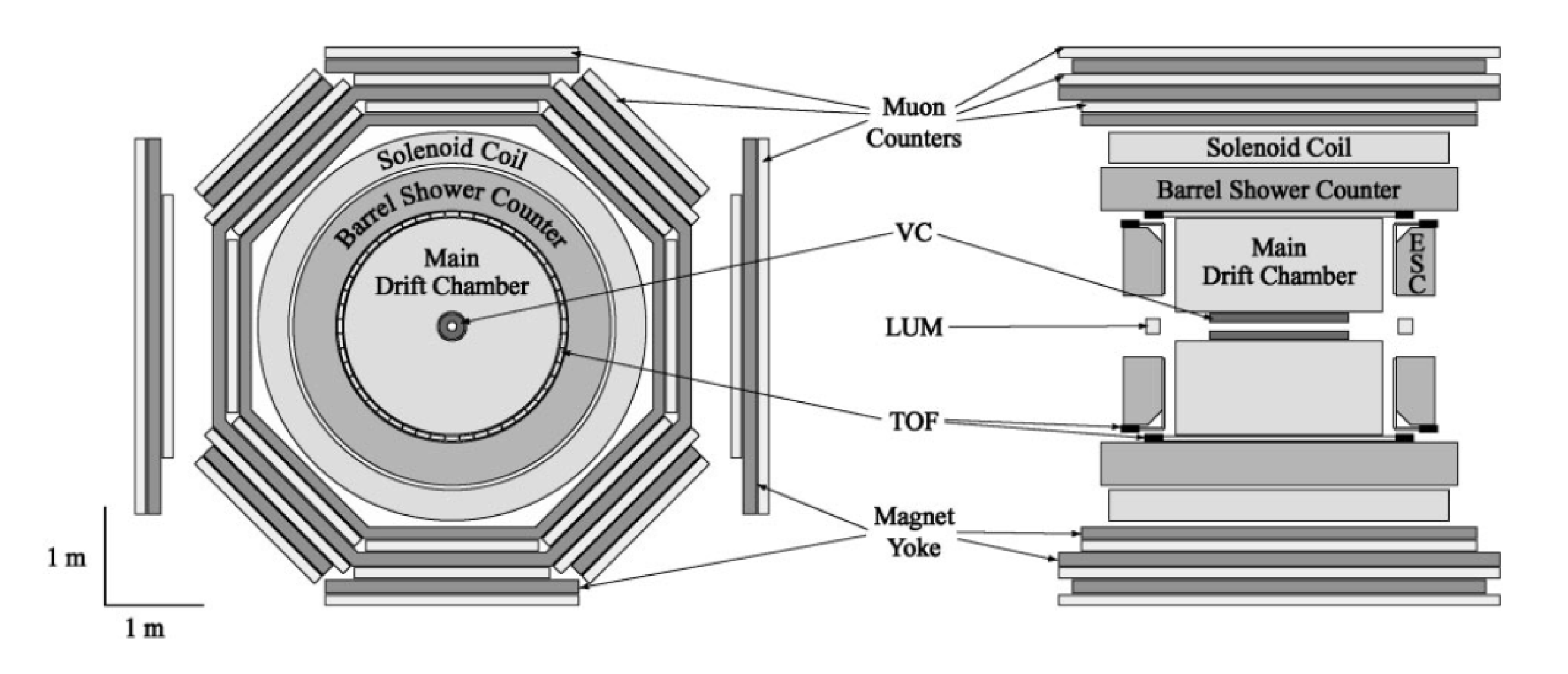}}
\caption{End view (left) and Sideview (right) of the BESII detector. \label{bes-dec}}
\end{figure}

The Beijing Spectrometer (BES) is a large general purpose solenoidal
detector\cite{bes1} at the Beijing Electron Positron Collider (BEPC), which
was designed for the $\tau$-charm physics in the center-of-mass energy range
of $2-5$ GeV.  The principal sub-detectors of BES are the central drift
charmber (CDC), the main drift chamber (MDC), the barrel and endcap shower
counters (BSC, ESC), magnet with a 0.4 T magnetic field, moun counters(MUC)
and luminosity monitor (LUMI).  Since its completion in 1989, the BES
detector had been in operation successfully.  About 9 million $J/\psi$ events,
4 million $\psi(2S)$ events and 22.3 nb$^{-1}$ data at 4.03 GeV were
collected for studying charmonium and charm decays.  Of particular
importance was a scan of the beam energy across the $\tau$ pair production
threshold, which was performed to precisely measure the $\tau$ mass. After
running for 6 years, aging effects were seen.  Thus a major upgrade on the
BES detector (called BESII afterwards)~\cite{bes2}, as displayed in
Fig.~\ref{bes-dec}, was made from 1996 to improve its performance.  Meanwhile,
the improvements on BEPC to increase the luminosity were performed, which are
described in Ref.\refcite{bepc}.  In addition, a GEANT3 based Monte Carlo
(MC) package (SIMBES)\cite{bes-sim}  with detailed consideration of the
BESII detector performance was used to improve the consistency between data
and MC.

The BESII detector started to take data in 1998. Since then a series of
important results, $e.g.$, precision $R$ value measurement\cite{rvalue-bes-2},
observation of X(1835)\cite{bes-x1835} , had been reported based on the
data samples at 6+85 center-of-mass energies between 2 and 5 GeV, $5.8\times
10^7$ $J/\psi$ events, $1.4\times 10^7$ $\psi(2S)$ events and 20 pb$^{-1}$
data at the peak of $\psi(3770)$, which underlined the rich physics in the
$\tau$-charm region, including light hadron spectroscopy, charmonium
spectrum, charm meson decays and $\tau$ physics.

The BES experiment is a unique facility for carrying out broad and
significant research in $\tau$-charm physics.  However, in this paper we
only focus on the two precision measurements, $\tau$ mass at BES and $R$
value at BESII, which are both very important in the test of the
Standard Model (SM).  The former one provides a significant test of lepton
universality, while the latter one is used to determine the vacuum
polarization, which plays a crucial role in the precision determination of
the QED running coupling constant evaluated at the $Z$ mass, $\alpha(M^2_Z)$,
and the anomalous magnetic moment of the muon, $a_\mu$.  The purpose of this
review is to look back at these two precision measurements, pointing out the research
in the energy region in $2-5$ GeV has been extraordinarily productive and
the exploration of this interesting and important physics has not even been
exhausted in the past.  Results relevant to this energy range have been, and
continue to be accumulated by the BESIII\cite{bes3-dec} experiment at BEPCII
(the only $e^+e^-$ collider in the world operating in this energy region).

\section{$\tau$ mass measurement at BES}	

$\tau$ lepton~\cite{dis-tau,distau2}, the most massive member of the charged lepton
family, was a giant step toward establishing the present SM.
In the past forty years, the $\tau$ lepton was a subject of extensive
experimental study since it offers a unique place to test fundamental
aspects of electroweak interactions.
Within the frame of SM the $e$, $\mu$ and $\tau$ are all point particles and
they have the same spin, electromagnetic and weak interactions except for
the different masses.  We usually refer to this as the lepton universality,
which could be tested by comparing the reaction with the interactions of
electrons and muons with the weak neutral current.  With an assumption of
$m_{\nu_\tau}=0$, the leptonic width in $\tau$ decays\cite{taudecay} is
given by,
\begin{equation}
\Gamma(\tau\rightarrow l \bar{\nu_l}\nu_\tau)= \frac{g^2_\tau g^2_l}{64m^4_W}\frac{m^5_\tau}{96\pi^3}f(\frac{m^2_l}{m^2_\tau})
(1+\frac{3m^2_\tau}{5m^2_W})[1+\frac{\alpha(m_\tau)}{2\pi}(\frac{25}{4}-\pi^2) ],
\end{equation}
where $f(x)=1-8x+8x^3-x^4-12x^2lnx$.

By comparing the partial widths for the leptonic decay modes the following
universality relation for the relative strengths of the $\tau$ and $\mu$
couplings is obtained,
\begin{equation}
\frac{g^2_\tau}{g^2_\mu}= \frac{m^5_\mu}{m^5_\tau} \frac{B(\tau\rightarrow e \bar{\nu_e}\nu_\tau)}{B(\mu\rightarrow e \bar{\nu_e}\nu_\mu)} \frac{\tau_\mu}{\tau_\tau},
\end{equation}
where the small radiative and electroweak corrections at a level of 0.0004
is neglected.

The world averages\cite{pdg92} in 1992 for the above quantities yielded
$\frac{g^2_\tau}{g^2_\mu}=0.941 \pm 0.025$, implying a 2.4 standard
deviation disagreement with lepton universality.  The above comparison
suggested rather strongly that significant shifts in $\tau_\tau$ and/or
$m_\tau$ should occur as new measurements became more precise.  However, the
consistency of many $\tau$ lifetime measurements over the years did not yet
indicate such a shift.  While the $\tau$ mass had not been well measured
before 1992.

DASP\cite{dasp} was the first experiment to use the energy dependence of the
$\tau^+\tau^-$ production cross section to measure the $\tau$ mass,
$m_\tau=1807\pm20$ MeV/c$^2$.  Later measurements by SPEC\cite{spec} and
DELCO\cite{dlco} were reported with results of $1787^{+10}_{-18}$ MeV/c$^2$
and $1783^{+3}_{-4}$ MeV/c$^2$, respectively.  Subsequently
MarkII\cite{mark2} present an indirect measurement of $1787\pm10$ MeV/c$^2$
by performing a simultaneous fit of the center of mass energy dependence of
the $\tau^+\tau^-$ production cross section and the pion energy spectrum in
the decay $\tau\rightarrow \pi \nu_\tau$.  The compilation of the above
results gave a weighted average value $m_\tau = 1784.1^{+2.7}_{-3.6}$
MeV/c$^2$\cite{pdg92}.  Given the large uncertainty of $\tau$ mass, it would
appear more likely that $m_\tau$ would come down in case of the lepton
universality.


With the advantage of the simplicity of the initial state at the $e^+e^-$
experiment, a determination of $\tau$ mass at the BES experiment was then
really necessary by measuring the $\tau^+\tau^-$ threshold production.  This
threshold is the lowest energy at which $\tau$ production is possible and is
directly related to the $\tau$ mass via $E_{thres}$ = 2$m_\tau$, here $E_{thres}$ is for the $\tau^+\tau^-$ production threshold.  Therefore
this measurement is independent of the mass of the $\tau$ neutrino,
$m_{\nu_\tau}$.

According to the previous measurements, the minimum threshold energy for producing $\tau$ lepton pairs is about 3.56 GeV. The best energy region for $\tau$ physics research at BES experiment is
 between the $\tau$ lepton pair threshold energy and 3.67 GeV, an energy just below the $\psi(2S)$ resonance.
In this energy region the only contamination of $\tau$ pair events comes from ordinary meson
production and well understood purely electromagnetic processes.
Above 3.67 GeV backgrounds from decays of charmed states grow rapidly.
Another important advantage of the $\tau$-charm region is the existence of two precisely known
narrow resonances, $J/\psi$ and $\psi(2S)$, which provide a very high rate signal to calibrate and monitor
the detector performance. With above considerations, the BES experiment is a unique facility to precise measure the $\tau$ mass, which allows for a tight control of the systematic uncertainties.

Most $\tau$ decays ($\sim$85\%) contains only one charged particle and at
least one neutrino which leaves no trace in the detector.  Clearly the
decays to $e$'s, $\mu$'s, $\pi$'s and $\rho$'s contain only one charged
particle and thus $\tau^+\tau^-$ production would be most prominent in
events with only two charged particles.  It is clear that $\tau^+\tau^-$
production can be most easily measured by studying $e^+e^-$ annihilation
events with two charged particles, which have the possibilities of $ee$,
$e\pi$, $\mu\mu$, $\mu h$ and $hh$ ($h$ represents for a charged $\pi$ or
$K$).  In order to maximize the signal-to-background ratio, these modes,
which have certain characteristics, making them more easily distinguishable
from background than others, could be used to measure $\tau^+\tau^-$
production.  From these considerations, both the decay modes,
$\tau^+\rightarrow e^+\nu_e\bar{\nu_\tau}$ and $\tau^-\rightarrow e^-
\bar{\nu_e} \nu_\tau$, which is referred as $e\mu$ mode, could be used to
tag $\tau^+\tau^-$ events.  By means of the information of charged particles
in sub-detectors of MDC, BSC/ESC and MUC, the $e\mu$ events could be well
distinguished from the background events, $e.g.,$ Bhabha, Dimuon, and
hadronic events.

To determine the $\tau$ mass, a likelihood-driven approach is proposed to
measure the $\tau$ pair production threshold energy.  According to the
previous measurements, the minimum energy for producing $\tau$ lepton pairs
is around 3.56 GeV.  In order to measure this energy we will take data at
various energies $E_i$.  At each energy we record: (1) the integrated
luminosity $\mathcal{L}_i$ accumulated at that energy, and ( 2) the number
of $\tau^+\tau^-$ events detected during that running period $n_i$.  At the
end of the search all of this data is fitted to a parameterized form for the
cross section via a likelihood function.  The value of the $\tau$ mass
parameter at the peak of the likelihood corresponds to the best estimate for
$m_\tau$.

We used a likelihood method during the search to estimate the most efficient
energy at which to run next, $i.e.$, $E_{i+1}$, depends on the set of data
accumulated thus far, $\{\{E_1,\mathcal{L}_1, n_1\}, ... , \{E_i,
\mathcal{L}_i, n_i\}\}$.  The full algorithm is described roughly below.

Given $N$ search points each of which has a number of $\tau^+\tau^-$ events
detected, a centre of mass energy $E$, and an integrated luminosity
$\mathcal{L}$, one may generate a likelihood $L$ as a product of
probabilities as follows,

\begin{equation}
L(m_\tau,\epsilon,\sigma_B)=  \prod_{i=1}^{}\frac{\mu^{N_i}_ie^{-\mu_i}}{N_i!},
\end{equation}
where $N_i$ is the number of observed $\tau^+\tau^-$ events at the $i$-th scan point; $\mu_i$ is the corresponding number of events expected, which is given by

\begin{equation}
\mu_i= [\epsilon r_i \sigma (W_i, m_\tau)+\sigma_B ]\mathcal{L_i},
\end{equation}

where  $m_\tau$ is the mass of the $\tau$ lepton, and $\epsilon$ is the
overall efficiency for identifying $\tau^+\tau^-$ events, which includes
branching fractions, trigger efficiency and detector efficiency; the
detector efficiency is obtained from the MC data surviving the $\tau$ event
selection criteria; $\sigma_B$ is an effective background cross section,
which assumed constant over the limited range of center of mass energy;
$W_i$ corresponds to the center-of-mass energy; $\mathcal{L}_i$ is the
integrated luminosity at scan point $i$, and $\sigma (W_i, m_\tau)$ is the
corresponding cross section for $\tau^+\tau^-$ production corrected for
Coulomb interaction, initial and final state radiations, vacuum
polarizations, and beam energy spread, which is given by

\begin{equation}
\sigma (W,m_\tau)= \frac{1}{\sqrt{2\pi}\Delta} \int^\infty_{2m_\tau} exp (\frac{-(W-W^\prime)^2}{2\Delta^2})dW^\prime
\int^{1-\frac{4m^2_\tau}{W^{\prime 2}}}_{0} F(x, W^\prime)\sigma_1(W^\prime\sqrt{1-x}, m_\tau)dx,
\end{equation}
where $\Delta$ is the energy spread, $F(x, W)$\cite{init-corr} is the
initial state radiation correction, and $\sigma_1$ is the cross
section\cite{voloshin}, $\sigma_1(W,
m_\tau)=\frac{4\pi\alpha^2}{3W^2}\frac{\beta(3-\beta^2)}{2}\frac{F_c(\beta)F_r(\beta)}{(1-\prod(W))^2}$.
Here $\beta=\sqrt{1-(\frac{2 m_\tau}{W})^2}$; $F_c(\beta)$ and $F_r(\beta)$
are the functions for the Coulomb interaction and final state radiation
corrections\cite{voloshin}; $\prod (W)$ is the vacuum polarization
corrections\cite{vacu-corr}.

In practice one would like to choose the set of search energies so that
we have some energies below threshold and some above.  This gives a more
accurate estimate of the $\tau$ mass, but it does mean that at some of our
search points is expected to observe only background events, if any.  In
principle, $\epsilon$ and $\sigma_B$ are allowed to vary along with $m_\tau$
in a three-dimensional likelihood fit to better estimate the systematic
errors due to those quantities.  But care must be taken to avoid trying to
fit too many parameters with too little data.  In our experiment we did not
utilize a multi-dimensional likelihood until after the search was completed.
During the energy scan, $\sigma_B$ could be kept fixed at a level determined
from studying background at the $J/\psi$ resonance; $\epsilon$ is kept fixed
at our best estimate from Monte Carlo studies.  The best estimate for
$m_\tau$ at any given time is the peak of the likelihood $L$ as a function
of $m_\tau$ only, using all information available up to that point.  The
next search energy is then derived from the likelihood by calculating the
energy most sensitive to the $\tau$ mass.

After the scan for each energy point, the $\tau$ pair production cross
section as a function of the center-of-mass energy is obtained from the
number of observed $\tau^+\tau^-$ event.  Then a maximum likelihood fit is
performed to identify the energy most sensitive the $\tau$ mass where the
$\tau$ mass is estimated from the peak of the likelihood.

In view of the importance of $\tau$ mass measurement and the unique
opportunity to determine the $\tau$ mass with greatly improved precision at
BES experiment, a proposal to measure the $\tau^+\tau^-$ production cross
section in the region most sensitive to the $m_\tau$ mass (just a few MeV
around threshold) was submitted to the National Laboratory of BEPC.  After
several rounds of the extensive discussions, this proposal was approved in
October, 1991.


Based on the above proposal, the selection of the experiment energy point is
crucial to the precision of the $\tau$ mass.  If these points are far above
the $\tau$ production threshold, as in the case of the previous experiments,
it is hard to improve the measurement precision on $\tau$ mass.  Therefore,
to measure the $\tau$ mass with great precision, of importance is to take
data just close to the $\tau$ pair production threshold.  Using the clean
$e\mu$ events as the signature of the $\tau$ pair production process,
$e^+e^-\rightarrow\tau^+\tau^-$, $\tau^-\rightarrow \mu^- (e^-)
\bar{\nu_\mu} (\bar{\nu_e})\nu_\tau$, $\tau^+\rightarrow \mu^+ (e^+)\nu_\mu
(\nu_e)\bar{\nu_\tau}$, we then select the energy points in the following
way.

To get the $\tau$ pair production threshold as soon as possible, we take the
$\tau$ mass, $m_\tau=1784$ MeV/c$^2$ (the world average in 1992), as the
single beam energy (the first energy scan point).  The observation of 2
$e\mu$ events in the data sample of 245.8 nb$^{-1}$ indicates that the the
$\tau$ pair production threshold is really below 1784 GeV/c$^2$.  After
tuning down 3 MeV, $E_2=1781$ MeV, we take an integrated luminosity of 248.9
nb$^{-1}$ data and found one $e\mu$ events.  As the third energy point reaches
$E_3=1772$ MeV, no $e\mu$ event is seen for the 232.8 nb$^{-1}$ data, which
implies that the $\tau$ pair production threshold above the $E_3$.
\begin{figure}[htpb]
\centerline{\includegraphics[width=8.8cm]{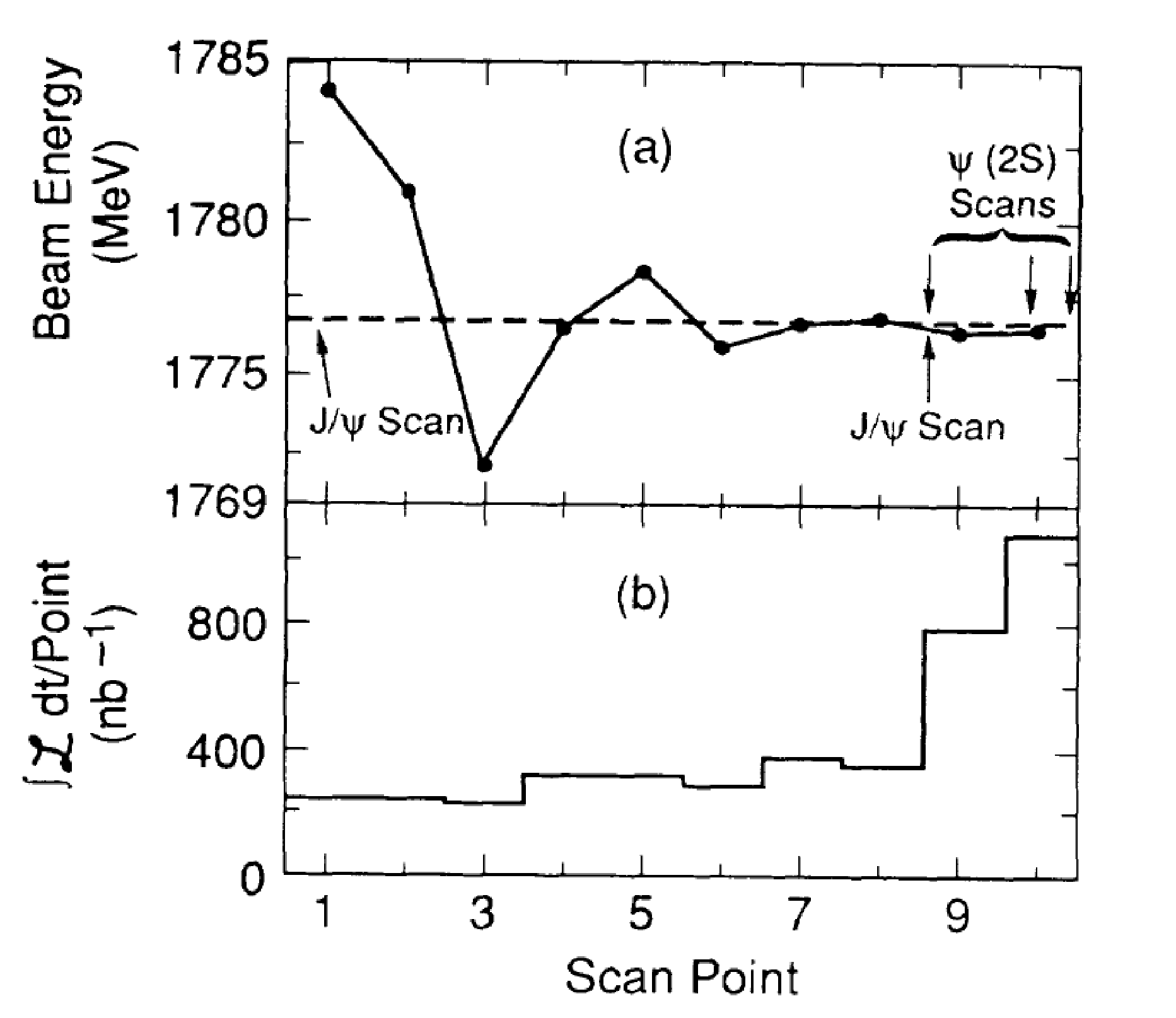}}
\caption{(a) The variation of the beam energy value with scan point;
(b) the integrated luminosity accumulated at each scan point.
\label{scan}}
\end{figure}

\begin{figure}[htpb]
\centerline{\includegraphics[width=8.8cm]{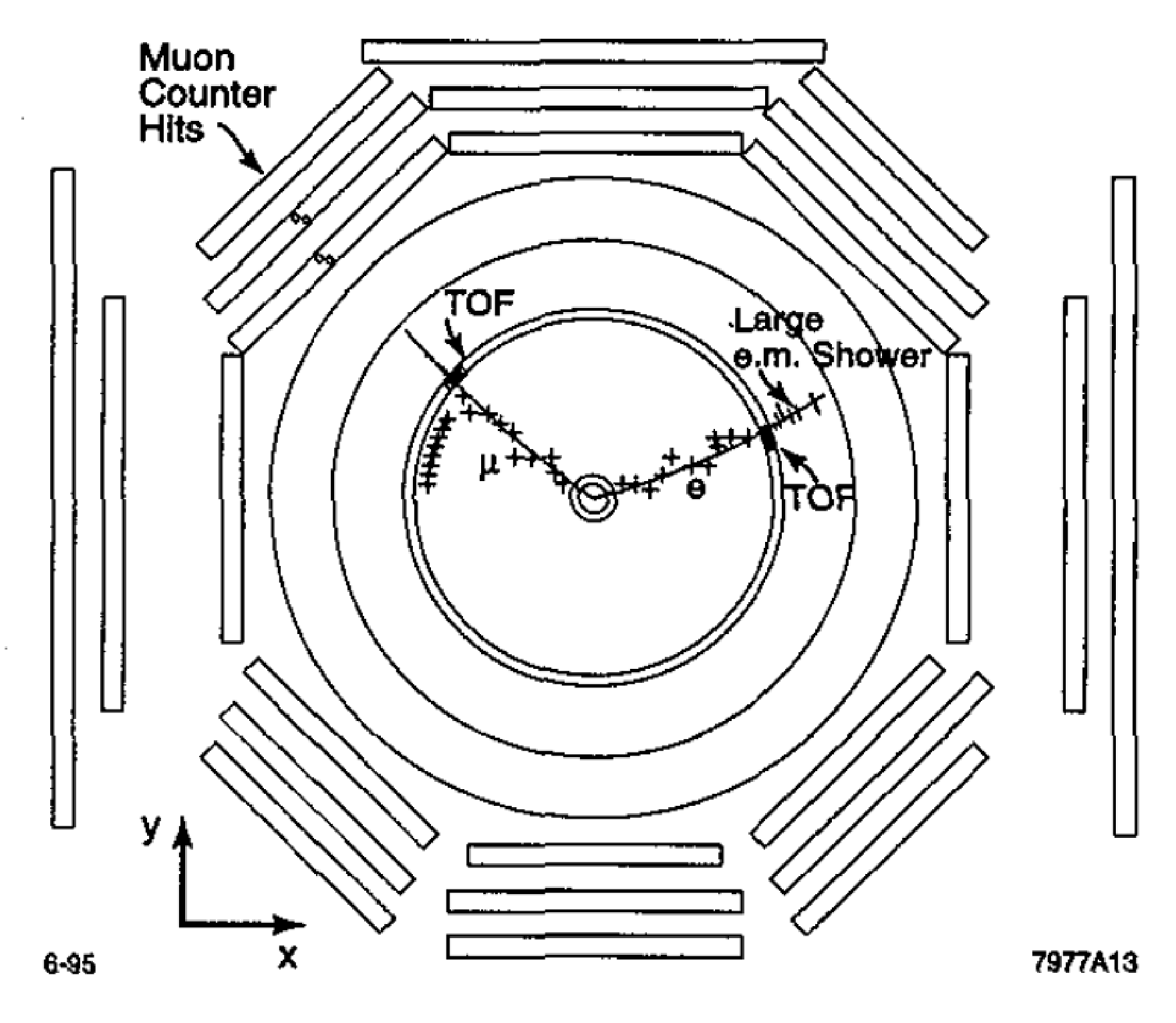}}
\caption{ A typical $e\mu$ event in $x-y$ projection.
 \label{96-event}}
\end{figure}

Following this strategy, 7 more energy points were scanned.  Thus the data
taking for 10 energy points were performed to determine to search for the
$\tau$ pair threshold within a range of 24 MeV.  A total of 5 pb$^{-1}$ of
$e^+e^-$ collision data were collected over a 5 month period beginning in
November 1991.  The sequence of each energy point and the corresponding
integrated luminosity are shown in Fig.~\ref{scan}.  To estimate the
detection efficiency, the additional two energy points were taken above the
threshold, where the cross section is relatively large and slowly varying
with the center of mass energy.  The detailed information for each energy
point are summarized in Table.~\ref{energyscan} and a typical event display
for one of the $e\mu$ event in $x-y$ projection is shown in
Fig.~\ref{96-event}.

\begin{figure}[htpb]
\centerline{\includegraphics[width=8.8cm]{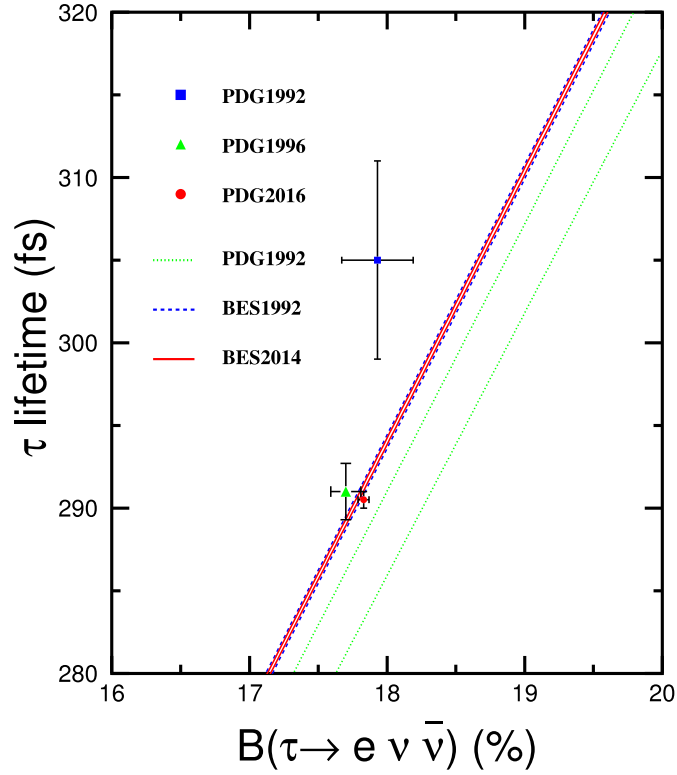}}
\caption{ The variation of $\tau_\tau$ with $B^e_\tau$ under the assumption of lepton universality; the one standard deviation bands calculated with $m_\tau$ form BES experiment (solid lines) and with that in 1992 PDG (dashed lines) are shown in comparison to the point corresponding to the world average values in 1992.
 \label{92-btau}}
\end{figure}

\begin{figure}[htpb]
\centerline{\includegraphics[width=5.8cm]{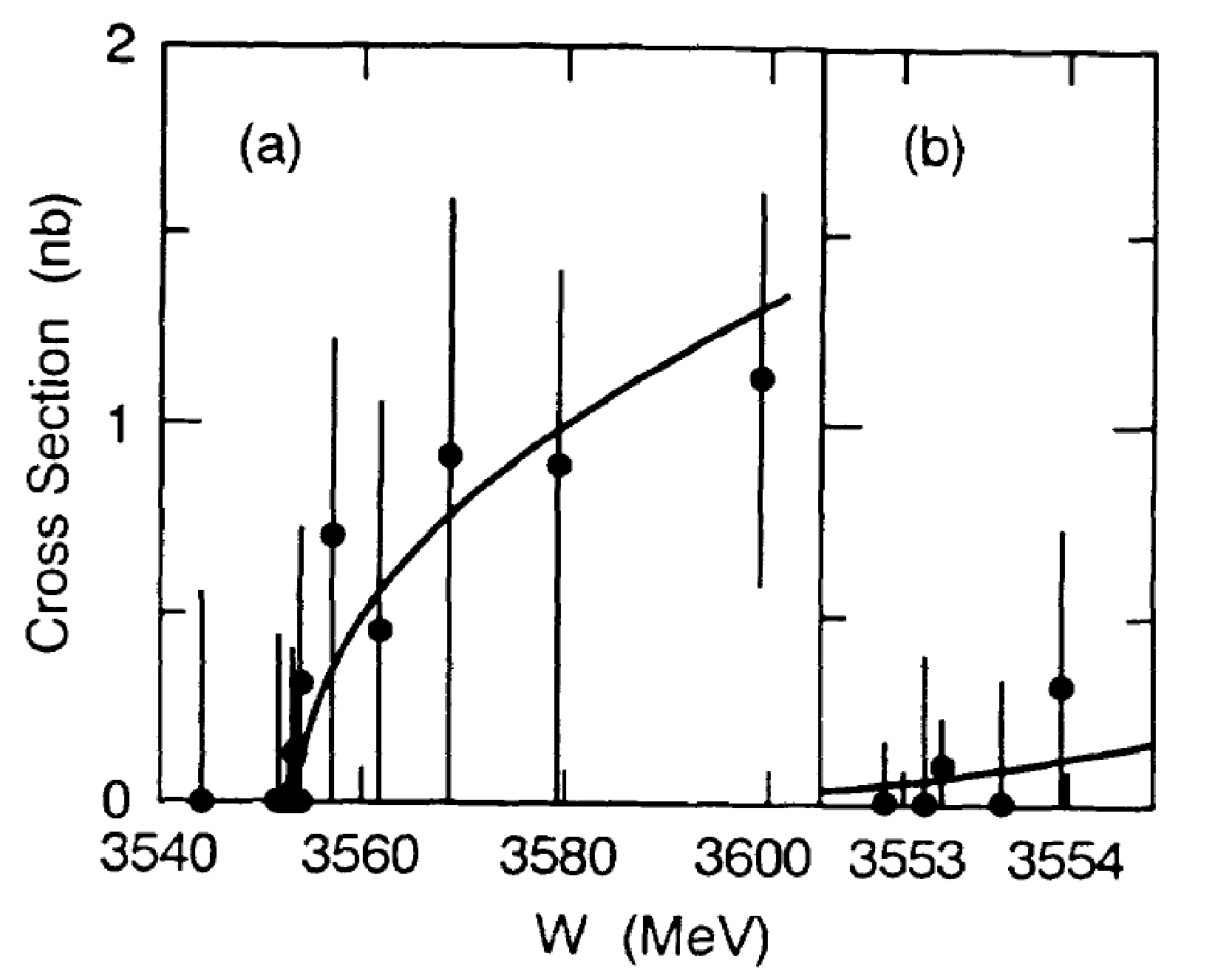}
\includegraphics[width=6.cm]{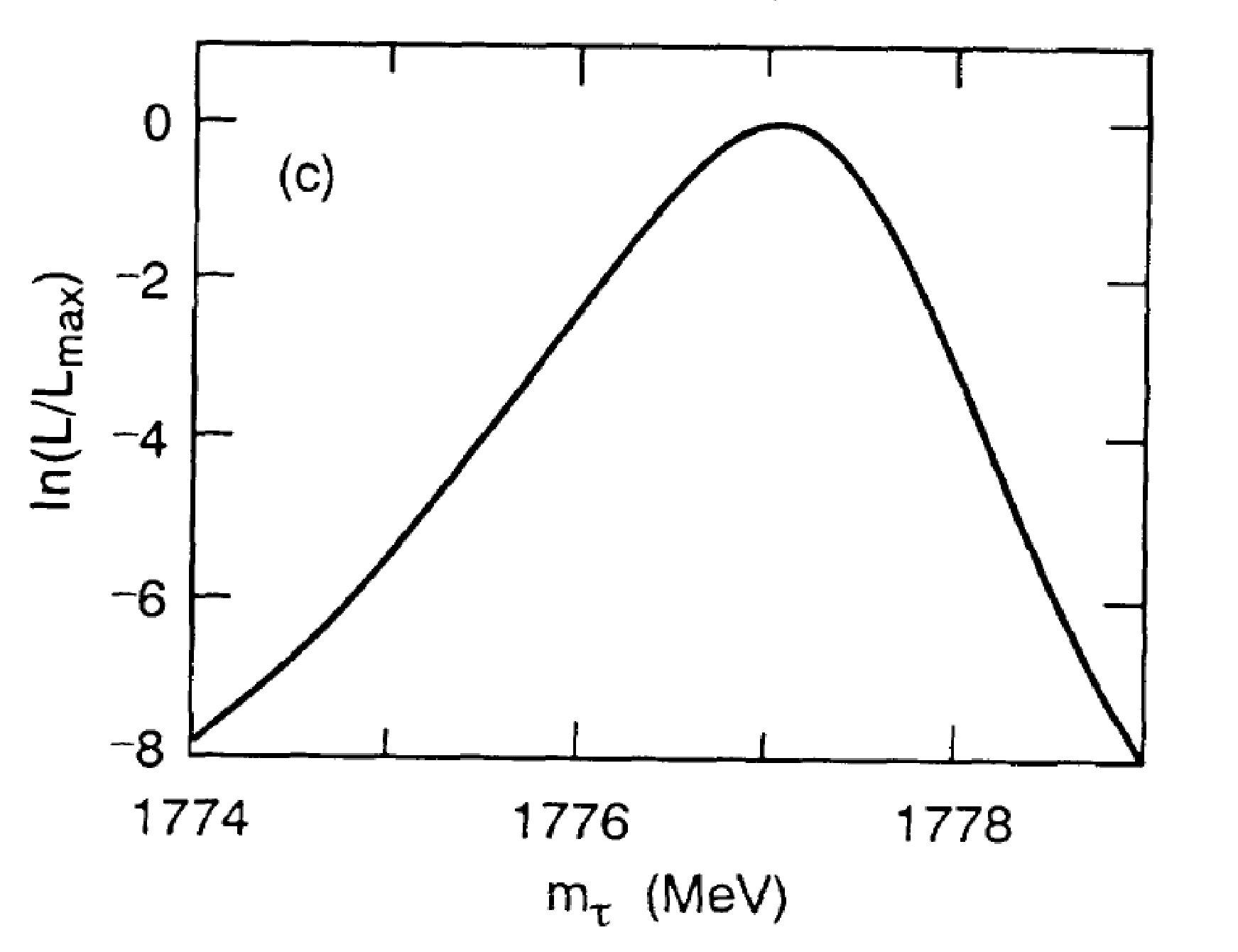}}
\caption{ (a) The center-of-mass energy dependence of the
cross section resulting from the likelihood fit;(b) An expanded version of (a),
in the immediate vicinity of $\tau^+\tau^-$ threshold. (c) The dependence
of the logarithm of the likelihood function on $m_\tau$ with
efficiency fixed at 14.1\%.
 \label{mtau}}
\end{figure}

After taking into account the center-of-mass energy spread, initial state
radiation and the vacuum polarization corrections, the efficiency corrected
cross sections as a function of corrected beam energy is shown in
Fig.~\ref{mtau} (a), which are described in detail in
Ref.\refcite{bes-tau1}.  In order to account for uncertainties in the
efficiency $\epsilon$, the branching fraction product and the luminosity.
The efficiency $\epsilon$ was treated as a free parameter in a
two-dimensional maximum-likelihood fit for $m_\tau$ and $\epsilon$ to the
data.  Finally the $\tau$ mass was determined to be
$m_\tau=1776.9^{+0.4}_{-0.5}\pm0.21$ MeV/c$^2$, which was 7.2 MeV (about two
standard deviation) below the world average value in 1992\cite{pdg92}, but
with significantly improved precision.  In this case, the coupling strength
ratio became $\frac{g^2_\tau}{g^2_\mu}=0.960 \pm 0.024$, The deviation from
lepton universality was reduced from 2.4 to 1.7 standard deviations as
indicated in Fig.~\ref{92-btau}.

The above measurement was performed with the $e\mu$ events only. To further
improve the measurement precision, the BES experiment\cite{bes-tau2}
reanalyzed the data by including additional decay modes $ee$, $e\pi$,
$\mu\mu$, $\mu h$ and $hh$, where $h$ represents for a charged $\pi$ or $K$,
and the $\tau$ mass was determined to be
$m_\tau=1776.9^{+0.18+0.25}_{-0.21-0.17}$ MeV/c$^2$.  With the latest
results of 1995\cite{othe-95}, $\tau_\tau=(291.6\pm 1.7)$ fs and
$B(\tau\rightarrow e\bar{\nu_e}\nu_\tau)=(17.66\pm0.11)\%$, the
$\frac{g^2_\tau}{g^2_\mu}=0.9886 \pm 0.0085$, which was consistent with
hypothesis of $\mu-\tau$ universality at a level of 1.3 standard deviation.

\begin{figure}[htpb]
\centerline{\includegraphics[width=8.8cm]{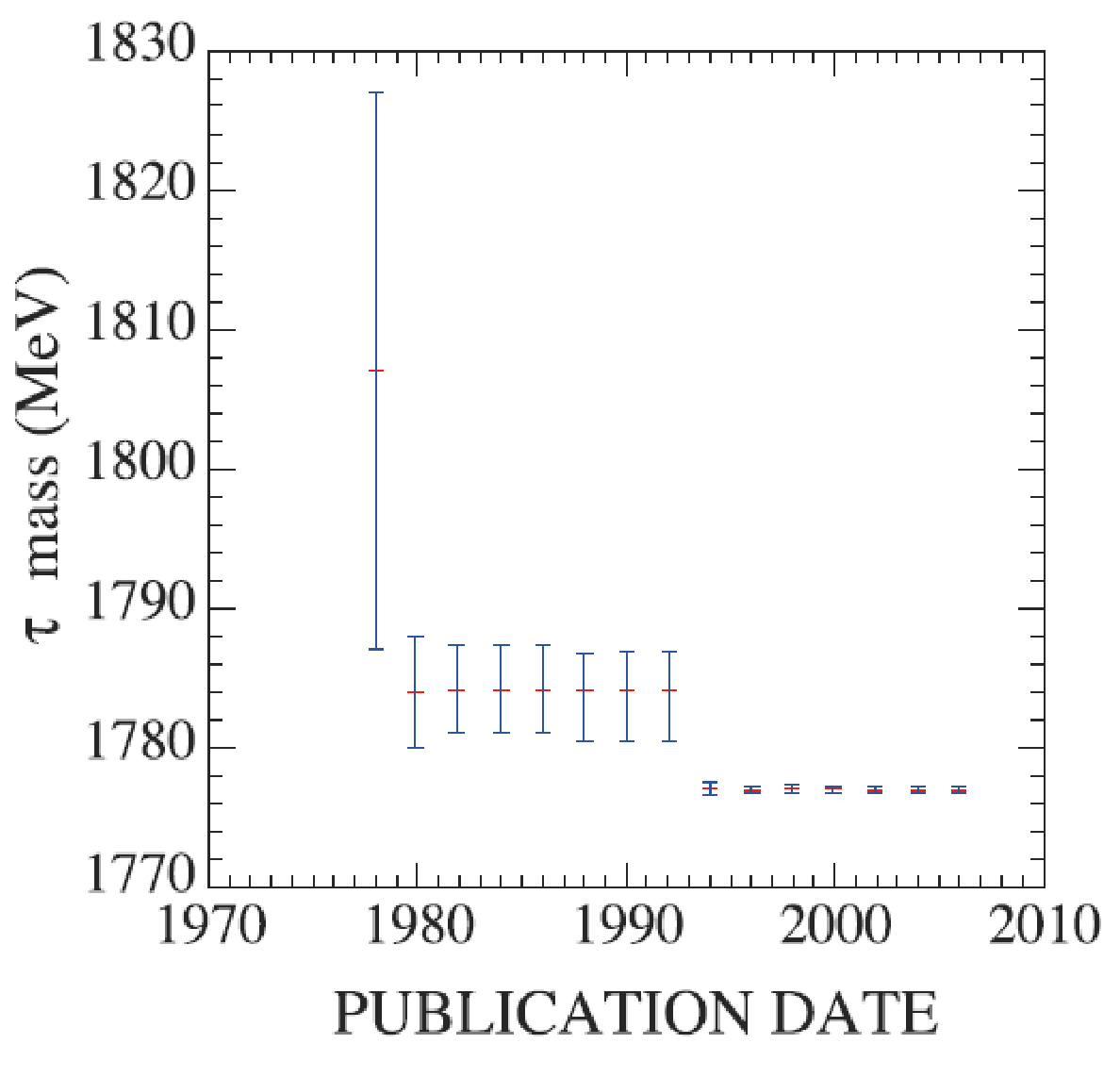}}
\caption{ The world average value of $\tau$ mass as a function of date of publication of PDG. This plot is from Ref.\cite{pdg06}
 \label{hist-mtau}}
\end{figure}

\begin{figure}[htpb]
\centerline{\includegraphics[width=8.8cm]{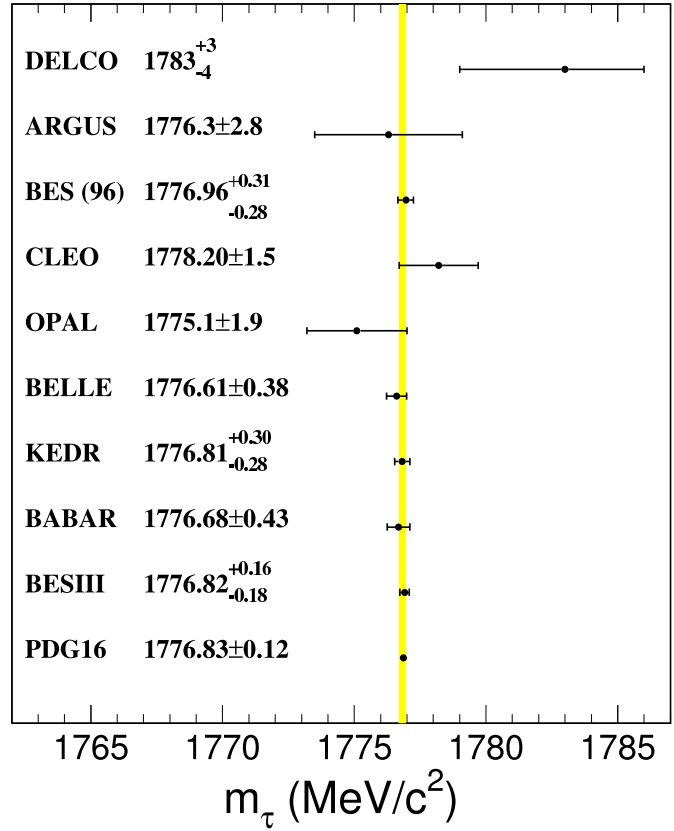}}
\caption{ Comparison of measured $\tau$ mass from
different experiments. The yellow band corresponds
to the 1$\sigma$ limit of the world average value.
 \label{mtaush}}
\end{figure}

Subsequently, a series results from different
experiments\cite{opal,kedr,belle,taumassbabar} confirmed this measurement, but the
average value is almost totally dominated by the BES measurement, which
could be clearly seen in Fig.~\ref{hist-mtau}.  Most recently the BESIII
experiment performed an energy scan around the $\tau$ pair production
threshold.  With the advantage of the beam energy measurement system
(BEMS)~\cite{bems}, which allows to determine the beam energy with high
precision, and the excellent performance of the BESIII detector, the $\tau$
mass is measured to be $m_\tau=1776.91\pm 0.12^{+0.10}_{-0.13}$
MeV/c$^2$,\cite{bes3-tau} which is the most precision measurement to date,
and the corresponding $\frac{g^2_\tau}{g^2_\mu}=1.0016\pm0.0042$, which is
in good agreement with the lepton universality as displayed in Fig.~\ref{92-btau}.  Of course, the precision
for the leptonic branching ratios and $\tau$ lifetime are also significantly
improved to be a level of 0.2\%.  The present situation on $\tau$ mass
measurement is summarized in Fig.\ref{mtaush}, where the vertical band
indicates the current world average value, $1776.83+\pm 0.12$
MeV/c$^2$.\cite{pdg16}.

\begin{table}[htpb]
\tbl{A chronological summary of the $\tau^+\tau^-$ threshold scan data; W denotes the corrected center-of-mass energy, $\Delta$ is the energy spread and $\mathcal{L}$ is the integrated luminosity.}
{\begin{tabular}{@{}ccccc@{}} \toprule
Scan point & W/2 (MeV) & $\Delta$ (MeV) & $\mathcal{L}$ (nb$^{-1}$) & N ($e\mu$ events)\\
\hline
1 &1784.19 &1.34& 245.8 &2 \\
2 &1780.99 &1.33 &248.9 &1\\
3 &1772.09 &1.36 &232.8 &0\\
4 &1776.57 &1.37 &323.0 &0\\
5 &1778.49 &1.44 &322.5 &2\\
6 &1775.95 &1.43 &296.9 &0\\
7 &1776.75 & 1.47&  384.0& 0\\
8 &1776.98 &1.47 &360.8 &1\\
9 &1776.45 &1.44 &794.1 &0\\
10& 1776.62& 1.40& 1109.1& 1\\
11&1799.51 &1.44 &499.7 &5\\
12 &1789.55& 1.43& 250.0 &2\\ \botrule
\end{tabular} \label{energyscan}}
\end{table}

\section{$R$ value measurement at BESII}


The $R$ is the ratio of the cross section of hadron production from the
annihilation of $e^+e^-$ into virtual $\gamma$ to that for muon pairs in the
lowest order,
\begin{equation}
R= \frac{\sigma (e^+e^-\rightarrow q\bar{q})}{\sigma(e^+e^-\rightarrow \mu^+\mu^-)},
\end{equation}
where the cross section of $e^+e^-\rightarrow \mu^+\mu^-$ is given by
$\sigma_{\mu^+\mu^-}=\frac{4\pi\alpha^2}{3s}$.  $\alpha$ is the fine
structure constant, $\sim1/137$, and $s$ is the squared centre-of-mass energy of
the $e^+e^-$ system.  Therefore the measurement of the values of $R$ is
eventually to measure the total cross section of the process
$e^+e^-\rightarrow hadrons$.

Originally the measurement of the energy dependence of the $R$ value was
used to test the existence of colored quarks within the naive quark
model, which provides fundamental reformation on the structure of hadrons.
At the tree-level, $R$ is given by $R= 3 \sum_{q} e^2_q$, where the factor
of 3 arises from the 3 quark colors, $e_q$ is the quark electric charge and
the summation is over all the quark flavours.  Without taking into account
the possible resonances directly produced from $e^+e^-$ annihilation, the
naive prediction for $R$ as a function of centre-of-mass energy is then a
constant with steps at the thresholds for quark pair production.  In
accordance with the early measurements from different
experiments~\cite{gamma2,mark1,pluto,pluto2} in the energy region from
$\pi\pi$ production threshold to the $Z^0$, the experimental $R$ values are
in general consistent with theoretical predictions, which confirmed the
hypothesis of the three color degrees of freedom for quarks.

\begin{figure}[htpb]
\centerline{\includegraphics[width=8.8cm]{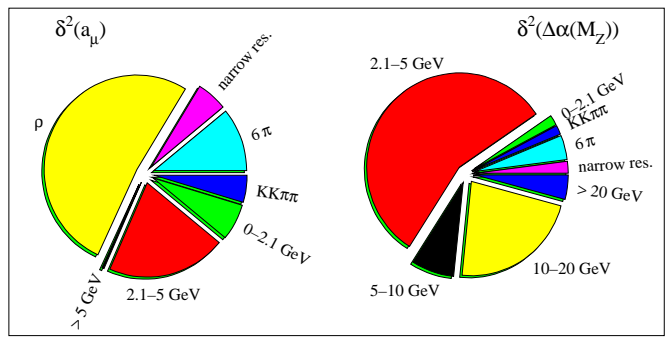}}
\caption{Quadratic contribution of the various error sources to
$\alpha_\mu$ (left hand plot) and $\alpha(M^2_Z)$(right hand plot) after
the inclusion of $\tau$ data. In the energy region $0-2.1$ GeV
we include all exclusive contributions that are not given
separately. Relative contributions to uncertainties of  $\alpha_\mu$
(left hand plot) and $\alpha(M^2_Z)$(right hand plot) from $R$ values
in different energy regions/resonances.
\label{alpha-err}}
\end{figure}

In addition, this quantity is also a necessary input for the experimental
evaluation of two important quantities for the precision test of the
Standard Model, which are $\alpha(M^2_Z)$, the electromagnetic coupling
constant evaluated at the mass of the $Z$ boson, and $\alpha_\mu$, the
anomalous magnetic moment of the muon.  Their theoretical precisions so far
are limited by the second order loop effects from the hadronic vacuum
polarization.  The uncertainties on $\alpha(M^2_Z)$ and $a_\mu$, as
indicated in Fig.~\ref{alpha-err}, are dominated by the contribution from
the $R$ value with uncertainties of $15\sim20$\% in the range of $1-5$
GeV~\cite{alpha-err}. Of importance is the $\alpha(M^2_Z)$ plays a vital role in the
determination of electroweak corrections relation to the mass of Higgs
particle, which was finally discovered in 2012~\cite{cms,atlas}, but as the only yet
undiscovered particle in the SM at that time.  An improved uncertainty for
$R$ value would help to determine $\alpha(s)$ and therefore result in
improved constraint on the parameters of the Standard Model, also contribute
to the interpretation of $\alpha_\mu$.

From Fig.~\ref{alpha-err}, the dominant uncertainty in $\alpha(M^2_Z)$ comes
from data in the energy range of $2-5$ GeV, thus making an improved
measurement is critical for better accuracy.  In the case of $\alpha_\mu$, a
smaller contribution is possible, due to the fact that $\alpha_\mu$ is more
sensitive to lower energy data. However, this contribution may become
important with the accumulation of lower energy data obtained from the
CMD and SND experiments at VEPP-2M in Novosibirsk, and KLOE experiment at DA$\Phi$NE in Frascati.

With the previous $e^+e^-$ experiments in the energy region of $2-5$ GeV, the
accuracy with which the absolute value of $R$ can be measured was limited by
systematic uncertainties.  In view of its importance, it is essential to
perform the $R$ value measurement at the BES experiment because the BEPC
just exactly operated in this energy region.  Before the official proposal,
we made a pre-study\cite{qixr} using the data, corresponding to an integrated
luminosity of 5 pb$^{-1}$, collected near $\tau^+\tau^-$ threshold in 1992
for the $\tau$ mass measurement.  The measured $R$ values around
center-of-mass energy of 3.55 GeV are in agreement with previous
measurements, but the precision was improved by a factor of 2,
which confirmed the feasibility of this proposal.

Just after the machine and detector upgrade~\cite{bes2}, the BES collaboration
performed two scans to measure $R$ in the energy region of $2-5$ GeV in 1998
and 1999.  To check the BESII detector performance and tune the
MC generator parameters for determining the detection efficiency of
inclusive hadronic events, first a test run scanned 6 energy points covering
the energy from 2.6 to 5 GeV in the continuum~\cite{rvalue-bes-1}.  The
integrated luminosity collected at each energy point changed from 85 to
292~nb$^{-1}$.  Separated beam running at each energy point was carried out
in order to subtract the beam associated background from the data.  After
that a fine scan of 85 energy points in the energy region of $2-4.8$
GeV\cite{rvalue-bes-2} was performed.  To subtract beam associated
background, separated beam running was done at 26 energy points and single
beam running for both $e^-$ and $e^+$ was done at 7 energy points
distributed over the whole scanned energy region.  Special runs were taken
at the $J/\psi$ to determine the trigger efficiency.  The $J/\psi$ and
$\psi(2S)$ resonances were also scanned at the beginning and at the end of
the $R$ scan for the energy calibration.

Due to the large number of final states, it is hard to determine the $R$
value by completely measuring all the hadronic processes.  The $R$ values at
BES experiment are measured by observing the final hadronic events
inclusively, i.e.  the value of $R$ is determined from the number of
observed hadronic events ($N^{obs}_{had}$), which is given by,
\begin{equation}
R=\frac{ N^{obs}_{had} - N_{bg} - \sum_{l}N_{ll} - N_{\gamma\gamma} }
{ \sigma^0_{\mu\mu} \cdot L \cdot \epsilon_{had} \cdot \epsilon_{trg}
\cdot (1+\delta)},
\end{equation}
where $N_{bg}$ is the number of beam associated background events;
$\sum_{l}N_{ll},~(l=e,\mu,\tau)$ and $N_{\gamma\gamma}$ are the numbers
of misidentified lepton-pairs from one-photon and two-photon processes
events; $L$ is the integrated luminosity; $\delta$ is the radiative
correction; $\epsilon_{had}$ is the detection efficiency for hadronic
events and $\epsilon_{trg}$ represents the trigger efficiency.

To distinguish the hadronic event from the single-photon production from all
other possible contamination mechanisms, a series of selection criteria were
applied by using all information of charged and neutral tracks from the
sub-detectors of BESII, including the requirements on the vertex position,
momentum of charged particle, the time of flight and the $\mu$ counter.
After that we performed an extensive study on the background contributions.
The dominant background events are from cosmic rays, lepton pair production
($e^+e^- \to e^+e^-$, $\mu^+\mu^-$, $\tau^+\tau^-$), two-photon processes,
and beam associated processes.  The cosmic rays could be easily removed by
using the TOF information.  The remaining background from lepton pair
production and two-photon processes is then subtracted out statistically
according to a MC simulation.

To estimate the beam associated background events, the none-colliding data
was taken at each energy point for the test run.  The salient features of
the beam associated background are that their tracks are very much along the
beam pipe direction, the energy deposited in BSC is small, and most of the
tracks are protons.  Therefore, most of them could be rejected by a vertex
cut.  With the same selection criteria as those for hadronic events, the
remaining background events in the hadronic events, after normaliziation in
accordance with the integrated beam currents for collision and separated
beam runs, were subtracted directly in the calculation of the hadronic cross
sections.  For the 1999 fine scan data, there were some improvements in
event selection and the numbers of hadronic events and beam-associated
background events were determined by fitting the distribution of event
vertices along the beam direction with a Gaussian to describe the hadronic
events and a polynomial of degree one to three for the beam-associated
background \cite{rvalue-bes-2}.

\begin{figure}[htpb]
\centerline{\includegraphics[width=8cm,height=6cm,]{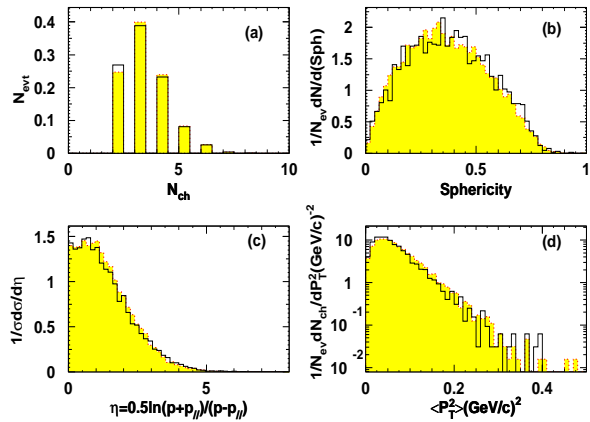}}
\caption{ Comparison of hadronic event shapes between data (shaded
region) and Monte Carlo (histogram).  (a) Multiplicity; (b) Sphericity;
(c) Rapidity; (d) Transverse momentum.
 \label{hadron-fig-1998}}
\end{figure}

The trigger efficiencies are measured by comparing the responses to
different trigger requirements in special runs taken at the $J/\psi$
resonance~\cite{trigger-bes}.  From the trigger measurements, the
efficiencies for Bhabha, dimuon and hadronic events are determined to be
99.96\%, 99.33\% and 99.76\%, respectively.  As a cross check, the trigger
information from the 2.6 and 3.55 GeV data samples is used to provide an
independent measurement of the trigger efficiencies.  This measurement is
consistent with the efficiencies determined from the $J/\psi$ data.  The
errors in the trigger efficiencies for Bhabha and hadronic events are less
than 0.5\%.

\begin{figure}[htpb]
\centerline{\includegraphics[width=8cm,height=6cm,]{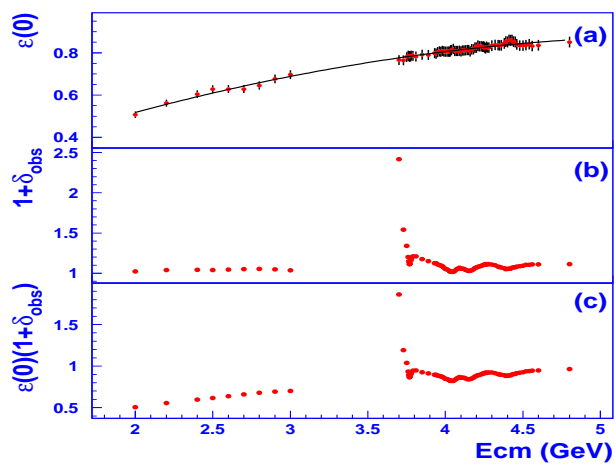}}
\caption{ (a) The c.m. energy dependence of the detection efficiency
for hadronic events estimated using the LUARLW generator. The error
bars are the total systematic errors.
(b) The calculated radiative correction, and
(c) the product of (a) and (b)~\cite{rvalue-bes-2}.
 \label{eff-fig-1999}}
\end{figure}

JETSET7.4 is used as the hadronic event generator to determine the
detection efficiency for hadronic events.  Parameters in the generator
are tuned using a $4 \times 10^4$ hadronic event sample collected near
3.55 GeV for the $\tau$ mass measurement done by the BES
Collaboration~\cite{bes-tau1,bes-tau2}. The parameters of the generator are
adjusted to reproduce distributions of kinematic variables such as
multiplicity, sphericity, transverse momentum, {\it etc}. However,
the Monte Carlo simulation packet JETSET was not designed to fully
describe few body states produced by $e^+e^-$ annihilation in the few
GeV energy region. A great effort has been made by the Lund group and
BES collaboration to develop the formalism using the basic Lund Model
area law directly for the Monte Carlo simulation, which is expected to
describe the data better~\cite{bo}. Figure~\ref{hadron-fig-1998} shows
the comparison of hadronic event shapes between data and MC simulations,
which indicates that the MC simulations could describe data well.
The detection efficiency curve with respect of the center-of-mass
energy is shown in Fig.~\ref{eff-fig-1999}.

Radiative corrections determined using four different
schemes~\cite{radcorr1,radcorr2,radcorr3,radcorr4} agreed with each
other within 1\% below charm threshold.  Above charm threshold, where
resonances are important, the agreement is around $1\sim3\%$.  The major
uncertainties common to all models are due to errors in previously
measured $R$-values and in the choice of values for the resonance
parameters. For the measurements reported here, we use the formalism of
Ref.~\refcite{radcorr3} and include the differences with the other schemes
in the systematic error of $2.2-4.1\%$.

\begin{figure}[htpb]
\centerline{\includegraphics[width=8cm,height=12cm,angle=-90]{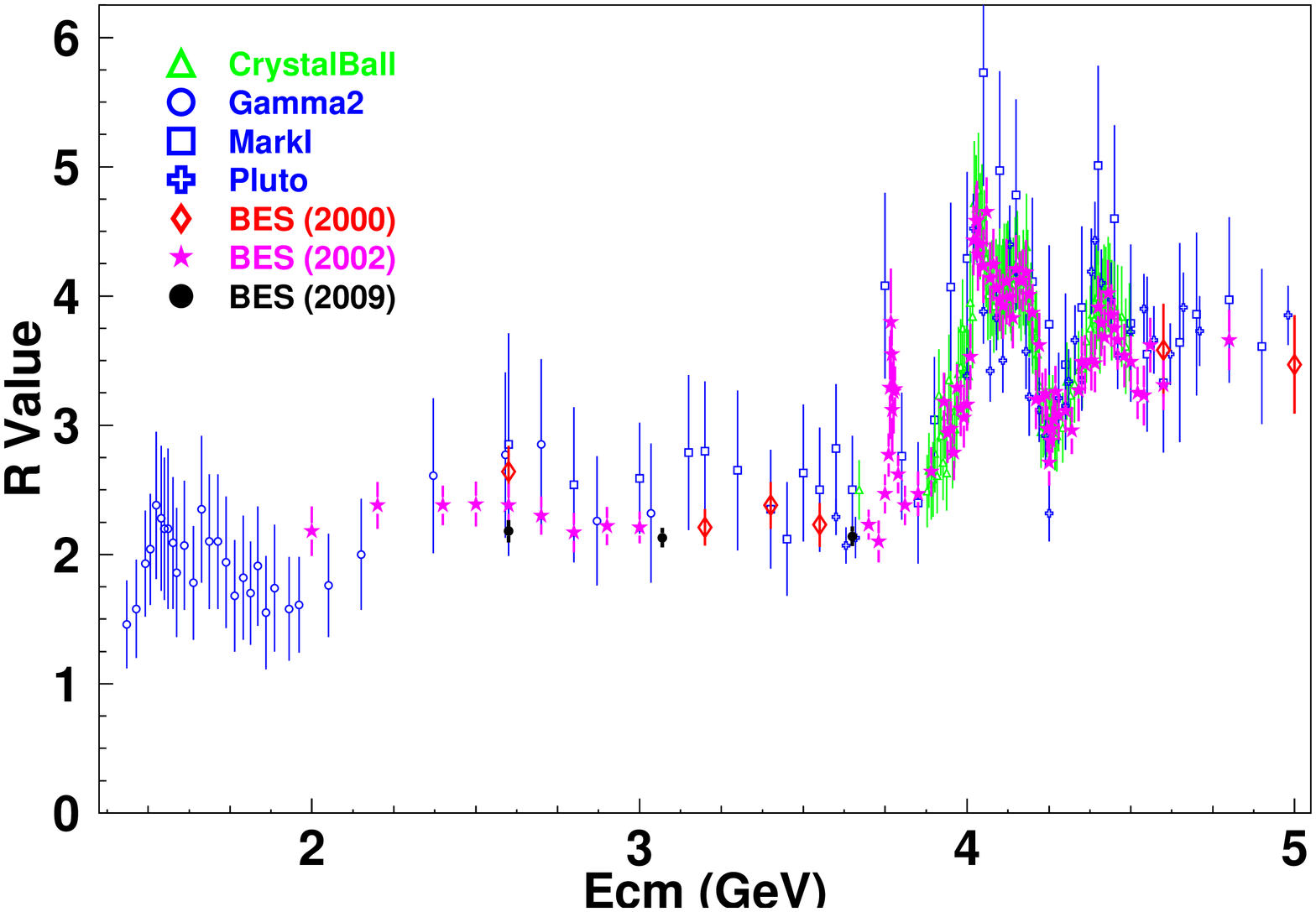}}
\caption{$R$ values at BES experiment together with previous
measurements below 5 GeV.
 \label{r-bes}}
\end{figure}

The $R$ values obtained at the 6 energy points scanned in 1998 and 85 energy
points in 1999 are summarized in Table~\ref{rvalue}, respectively, and
graphically displayed in Fig.~\ref{r-bes}, together with those measured by
MarkI, $\gamma\gamma 2$, and Pluto~\cite{mark1,gamma2,pluto}.  The $R$
values from BESII have an average uncertainty of about 6.6\%, which
represents a factor of two to three improvement in precision in the 2 to 5
GeV energy region.  Of this error, 3.3\% is common to all points.  These
improved measurements had a significant impact on the the global fit to the
electroweak data and the determination of the SM prediction for the mass of
the Higgs particle \cite{bolekpl}.  In addition, they provided an
improvement in the precision of the calculated value of
$a_{\mu}^{SM}$~\cite{martin}, and test the QCD sum rules down to 2
GeV~\cite{dave,kuehn}.

\begin{table}[htbp]
\tbl{Summary of $R$ data and values~\cite{rvalue-bes-1,rvalue-bes-2}.}
{\begin{tabular}{@{}cccccc@{}} \toprule
\hline
$E_{cm}$ & $R$   & $E_{cm}$ & $R$ &   $E_{cm}$ & $R$  \\ \hline
\hline
\multicolumn{6}{c}{BES 1998 data\cite{rvalue-bes-1}}\\
\hline
2.60 &$2.64\pm0.05\pm0.19$ &3.40 &$2.38\pm0.07\pm 0.16$ &4.60 &$3.58\pm0.20\pm 0.29$\\
3.20 &$2.21\pm 0.07\pm 0.13$ &3.55 &$2.23\pm0.06\pm0.16$& 5.00 &$3.47\pm0.32\pm 0.29$\\
\hline
\multicolumn{6}{c}{BES 1999 data\cite{rvalue-bes-2}}\\
\hline
2.000& $2.18\pm0.07\pm0.18$ & 3.990& $3.06\pm0.15\pm0.18$    & 4.245 & $2.97\pm0.11\pm0.14$ \\
2.200& $2.38\pm0.07\pm0.17$ & 4.000& $3.16\pm0.14\pm0.15$    & 4.250 & $2.71\pm0.12\pm0.13$  \\
2.400& $2.38\pm0.07\pm0.14$ & 4.010& $3.53\pm0.16\pm0.20$    & 4.255 & $2.88\pm0.11\pm0.14$\\
2.500& $2.39\pm0.08\pm0.15$ & 4.020& $4.43\pm0.16\pm0.21$    & 4.260 & $2.97\pm0.11\pm0.14$\\
2.600& $2.38\pm0.06\pm0.15$ & 4.027& $4.58\pm0.18\pm0.21$    &  4.270 & $3.26\pm0.12\pm0.16$  \\
2.700& $2.30\pm0.07\pm0.13$ & 4.027& $4.58\pm0.18\pm0.21$    &  4.280 & $3.08\pm0.12\pm0.15$  \\
2.800& $2.17\pm0.06\pm0.14$ & 4.030& $4.58\pm0.20\pm0.23$    & 4.300 & $3.11\pm0.12\pm0.12$\\
2.900& $2.22\pm0.07\pm0.13$ & 4.040& $4.40\pm0.17\pm0.19$    &  4.320 & $2.96\pm0.12\pm0.14$\\
3.000& $2.21\pm0.05\pm0.11$ &4.050& $4.23\pm0.17\pm0.22$     & 4.340& $3.27\pm0.15\pm0.18$ \\
3.700& $2.23\pm0.08\pm0.08$ &4.060& $4.65\pm0.19\pm0.19$     & 4.350& $3.49\pm0.14\pm0.14$\\
3.730& $2.10\pm0.08\pm0.14$ &4.070& $4.14\pm0.20\pm0.19$     & 4.360& $3.47\pm0.13\pm0.18$ \\
3.750& $2.47\pm0.09\pm0.12$ &4.080& $4.24\pm0.21\pm0.18$     & 4.380& $3.50\pm0.15\pm0.17$ \\
3.760& $2.77\pm0.11\pm0.13$ &4.090& $4.06\pm0.17\pm0.18$     & 4.390& $3.48\pm0.16\pm0.16$ \\
3.764& $3.29\pm0.27\pm0.29$ &4.100& $3.97\pm0.16\pm0.18$     & 4.400& $3.91\pm0.16\pm0.19$\\
3.768& $3.80\pm0.33\pm0.25$ &4.110& $3.92\pm0.16\pm0.19$     &4.410& $3.79\pm0.15\pm0.20$\\
3.770& $3.55\pm0.14\pm0.19$ &4.120 & $4.11\pm0.24\pm0.23$    & 4.420& $3.68\pm0.14\pm0.17$  \\
3.772& $3.12\pm0.24\pm0.23$ &4.130 & $3.99\pm0.15\pm0.17$    & 4.430& $4.02\pm0.16\pm0.20$\\
3.776& $3.26\pm0.26\pm0.19$ &4.140 & $3.83\pm0.15\pm0.18$    &4.440& $3.85\pm0.17\pm0.17$ \\
3.780& $3.28\pm0.12\pm0.12$ &4.150 & $4.21\pm0.18\pm0.19$    &4.450& $3.75\pm0.15\pm0.17$ \\
3.790& $2.62\pm0.11\pm0.10$ &4.160 & $4.12\pm0.15\pm0.16$    & 4.450& $3.75\pm0.15\pm0.17$                    \\
3.810& $2.38\pm0.10\pm0.12$ &4.170 & $4.12\pm0.15\pm0.19$    &  4.460& $3.66\pm0.17\pm0.16$                  \\
3.850& $2.47\pm0.11\pm0.13$ &4.180 & $4.18\pm0.17\pm0.18$    &4.480& $3.54\pm0.17\pm0.18$                      \\
3.890& $2.64\pm0.11\pm0.15$ &4.190 & $4.01\pm0.14\pm0.14$    &4.500& $3.49\pm0.14\pm0.15$  \\
3.930& $3.18\pm0.14\pm0.17$ &4.200 & $3.87\pm0.16\pm0.16$    &4.520& $3.25\pm0.13\pm0.15$ \\
3.940& $2.94\pm0.13\pm0.19$ & 4.210 & $3.20\pm0.16\pm0.17$   &4.540& $3.23\pm0.14\pm0.18$\\
3.950& $2.97\pm0.13\pm0.17$ &4.220 & $3.62\pm0.15\pm0.20$    &4.560& $3.62\pm0.13\pm0.16$\\
3.960& $2.79\pm0.12\pm0.17$ &4.230 & $3.21\pm0.13\pm0.15$ &   4.600& $3.31\pm0.11\pm0.16$ \\
3.970& $3.29\pm0.13\pm0.13$ & 4.240 & $3.24\pm0.12\pm0.15$&  4.800& $3.66\pm0.14\pm0.19$ \\
3.980& $3.13\pm0.14\pm0.16$ & & & & \\
\hline
\multicolumn{6}{c}{BES 2004 data~\cite{rvalue-bes-3}}\\
\hline
2.60 &$2.18\pm0.02\pm0.08$ &3.07 & $2.13\pm0.02\pm0.07$ &3.65 &  $2.14\pm0.01\pm0.07$\\
\hline
\end{tabular}
\label{rvalue}}
\end{table}

In 2004, before BESII was shut down for the upgrade to BESIII, a
high-statistics data sample was taken at 2.6, 3.07 and 3.65~GeV, with an
integrated luminosity of 1222, 2291 and 6485~nb$^{-1}$,
respectively~\cite{rvalue-bes-3}.  The results are summarized in
Table.~\ref{rvalue}.  Compared with the previous results from BESII experiment
\cite{rvalue-bes-1,rvalue-bes-2}, the measurement precision was
improved due to three main refinements to the analysis: (1) the simulation
of BESII included both of the hadronic and electromagnetic interactions with
a GEANT3\cite{GEANT3A} based package SIMBES\cite{bes-sim} with a
more detailed geometrical description and matter definition for the
sub-detectors; (2) large data samples were taken at each energy point, with
statistical errors smaller than $1\%$; (3) the selected hadronic event
sample was expanded to include one-track events, which supplied more
information to the tuning of LUARLW, and resulted in the improved values of
parameters and hadronic efficiency.  With these improvements, the errors on
the new measured $R$ values were reduced to about $3.5\%$.

Using the BESII measurements, together with the results from other
experiments\cite{pipiISRbabar,cmd2,kloe,kloe2,snd} around that time, the hadronic
contribution to the running of the QED fine structure constant was
evaluated to be $\alpha_{s}^{(5)}(M_Z^2) = 0.02750\pm0.00033$~\cite{alphaz},
so that the preferred Higgs mass value was increased from $89^{+35}_{-26}$
to $93^{+35}_{-27}$ GeV/c$^2$ and the one-sided 95\% confidence level
upper limit was from 158 to 163 GeV/c$^2$.

Besides the dedicated project for $R$ scans, BESII also published $R$
values from the data taken primarily for other physics programs. In the
study of non-$D\bar{D}$ decay of $\psi(3770)$, data samples at 3.65 GeV
with a luminosity of 5.536 pb$^{-1}$ and at 3.665 GeV with a luminosity
of 998.2 nb$^{-1}$ were collected, with which the hadronic events with
more than 2-tracks ($n_{ch}\geq3$) were selected, and the averaged
$R$ value was obtained as $R=2.218\pm0.019\pm0.089$ with an error of
$4.1\%$\cite{plb6412006145}.
For the measurement of the resonant parameters of $\psi(3770)$, a scan
with 68 energy points in the energy region between 3.650 and 3.872 GeV
was performed, and this data sample was used for the $R$ measurement
as well, with an overall systematic uncertainty of $4.0\sim4.9\%$
and $R_{uds} = 2.141 \pm 0.025 \pm 0.085$ for the continuum light hadron
production near the $D\bar{D}$ threshold \cite{r-psi3770}.

\section{Summary and Outlook}

In the past two decades, our knowledge of the $\tau$ properties has been
considerably improved and all experimental results
on the $\tau$ lepton are consistent with the SM.  By means of maximum
likelihood approach, the BES experiment performed a fine scan just around
the $\tau$ pair production threshold and presented a precise determination
of $\tau$ lepton mass, $m_\tau=1776.9^{+0.18+0.25}_{-0.21-0.17}\pm0.21$
MeV/c$^2$.  Then the coupling strength ratio $\frac{g^2_\tau}{g^2_\mu}$ was
calculated to be $0.9886 \pm 0.0085$, which confirmed the lepton
universality and therefore resolved the discrepancy observed in early 1990s.
With the excellence performance of BEMS, the BESIII experiment yielded the
most precision measurement to date, $m_\tau=1776.91\pm 0.12^{+0.10}_{-0.13}$
MeV/c$^2$.  In this case the lepton universality has been tested to rather
good accuracy with $\frac{g^2_\tau}{g^2_\mu}=1.0016\pm0.0042$.  However,
there is still large room for improvements by comparing the precision of the
masses of $e$ and $\mu$.  Future measurement with high precision will
therefore allow to probe the lepton universality to a much deeper level of
sensitivity.

In addition, being one of the most fundamental parameters in particle
physics, the $R$-value plays an important role in the development of the
theory of particle physics and in the test of SM.  Experimental
efforts to precisely measure $R$ values at low energies are crucial for the
future electroweak precision physics.  The BESII experiment performed
energy scans in the energy region between 2 and 5 GeV, and the $R$ values
were measured at 6+85 energy points with an average uncertainty of 6.6\%,
representing a significant improvement in precision by a factor of 2 or 3
and thus a great impact on the determination of $\alpha(M^2_Z)$.

However, the present uncertainty on the $R$ values in the low energy region
of $2-5$ GeV is still at a level of $\sim$6\%, which should be further
improved for the precision test of SM. The BESIII experiment is therefore
absolutely crucial for a better determination of the $R$ values in this
energy region. A test run at 4 energies, 2.2324, 2.4, 2.8 and 3.4 GeV,
with a total integrated luminosity about 12 pb$^{-1}$ was carried out in
2011. Together with other data samples in the continuum for various projects,
a feasibility study of $R$ measurement at BESIII has been performed which
shows a precision of $\sim3$\% can be reached. Then a dedicated scan was done
in the high energy region of $3.85-4.6$ GeV with $\sim800$ pb$^{-1}$ at
104 energy points and at least $100k$ hadronic events at each point in
the $2013-2014$ run, followed by a low energy scan in $2.0-3.08$ GeV with
$\sim525$ pb$^{-1}$ at 21 energy points in the $2014-2015$ run.
These data samples make it possible for an improvement by a factor
of 2 once again in the $R$ measurement.
Since the luminosity of BEPCII is two orders of magnitude higher
than at BEPC, the scan of the resonance region will provide precise
information on the $1^{--}$ charmonium states up to 4.6~GeV.  The analysis
is undergoing and the results will be presented in the near future, which
will be very important in electroweak theory physics with regard to the
$R$-values at low energy region.

In the past, the research in the $\tau-charm$ region has been
extraordinarily productive and among other discoveries has revealed the
existence of the $\tau$ lepton and of the bound and rare states of the charmed
quarks.  Since this energy region encompasses a rich spectroscopy of
charmonium states, the charm meson and $\tau$ physics, exploration of this
interesting and important physics in this energy region is far to be exhausted in the past.
Therefore $\tau-charm$ physics will continue to be studied at facilities,
such as BESIII, Belle-II.  The BESIII experiment will provide competitive experimental conditions for carrying out
significant physics in many areas of $\tau-charm$ physics.

\section*{Acknowledgments}
The authors thank the BES colleagues, the staff of BEPC and the computing center at the Institute of High
Energy Physics, Beijing, for their hard efforts. This work is supported in part by the National Natural Science
Foundation of China under contracts Nos. 11675184, 11335008 and 2015CB856705.

\appendix



\begin{thebibliography}{0}    


\bibitem{bes1} J. Z. Bai {\it et al.} (BES Collaboration), {\it  Nucl. Instrum. Meth. A\/} {\bf 344},319 (1994).
\bibitem{bes2} J. Z. Bai {\it et al.} (BES Collaboration), {\it  Nucl. Instrum. Meth. A\/} {\bf 458},627 (2001).
\bibitem{bepc} Y. Wu, Operational status and future upgrades of the
BEPC, Talk given at the 18th Particle Accelerator Conference, New York, March 1999.
\bibitem{bes-sim} M. Ablikim {\it et al.} (BES Collaboration), {\it  Nucl. Instrum. Meth. A\/} {\bf 552},344 (2005).
\bibitem{rvalue-bes-2} J. Z. Bai  {\it et al.} (BES Collaboration), {\it Phys. Rev. Lett.} {\bf 88}, 101802 (2002).
\bibitem{bes-x1835} M. Ablikim {\it et al.} (BES Collaboration), {\it  Phys. Rev. Lett.} {\bf 95},262001 (2005).
\bibitem{bes3-dec} M. Ablikim {\it et al.} (BESIII Collaboration), {\it  Nucl. Instrum. Meth. A\/} {\bf 614},345 (2010).

\bibitem{dis-tau} M. L. Perl {\it et al.} , {\it Phys. Rev. Lett.} {\bf 35}, 1489 (1975).
\bibitem{dis-tau2} M. L. Perl {\it et al.}, {\it Phys. Rev. Lett.} {\bf 38}, 117 (1976).
\bibitem{taudecay} W. Marciano,	 {\it Ann. Rev. Nucl. Part. Sci.} {\bf 41}, 469 (1991).
\bibitem{pdg92} K. Hikas {\it et al.}, {\it Phys. Rev. D} {\bf 45}, 1 (1992).
\bibitem{dasp} R. Brandelik {\it et al.} (DASP Collaboration), {\it Phys. Lett.} {\bf B73}, 109 (1978).
\bibitem{spec} W. Bartel {\it et al.} (SPEC Collaboration), {\it Phys. Lett.} {\bf B77}, 331 (1978).
\bibitem{dlco} W. W. Bacino {\it et al.} (DLCO Collaboration), {\it Phys. Rev. Lett.} {\bf 41},13 (1978).
 \bibitem{mark2} C.A. Blocker, Ph.D. Thesis, LBL-Report 10801 (1980).

 \bibitem{init-corr} E. A. Kuraev and V. S. Fadin, {\it Yad. Fiz.} {\bf 41}, 733(1985).
 \bibitem{voloshin} M. B. Voloshin, {\it Phys. Lett. B} {\bf 556}, 153 (2003).
 \bibitem{vacu-corr} F. A. Berends and G. J. Komen, {\it Phys. Lett. } {\bf B63}, 432 (1976)
\bibitem{bes-tau1} J. Z. Bai {\it et al.} (BES Collaboration), {\it Phys. Rev. lett.} {\bf 69}, 3021(1992).
\bibitem{bes-tau2} J. Z. Bai {\it et al.} (BES Collaboration), {\it Phys. Rev. lett. D} {\bf 53}, 20(1996).
\bibitem{othe-95}  J. R. Patterson, in Proceedings of the XXVII-th International Conference on High Energy Physics, Glasgow,
Scotland, 1994, edited by P. J. Bussey and I. G. Knowles (IOP, London, 1995), Vol. I, p. 149.
\bibitem{opal} G. Abbiendi {\it et al.} (OPAL Collaboration), {\it Phys. Lett. B} {\bf 492}, 23, (2000).
\bibitem{kedr} V. V. Anashin {\it et al.} (KEDR Collaboration), {\it JETP Lett.} {\bf 85}, 347(2007).
\bibitem{belle} K. Belous {\it et al.} (BELLE Collaboration), {\it Phys. Rev. Lett.} {\bf 99}, 011801(2007).
\bibitem{taumassbabar} B. Aubert {\it et al.} (Babar Collaboration), {\it Phys. Rev. D} {\bf 80}, 092005(2009).
\bibitem{pdg06}  W. M. Yao {\it et al.} (Particle Data Group), {\it J. Phys. G} {\bf 33}, 1 (2006).
\bibitem{bems}  E. V. Abakumova {\it et al.}, {\it Nucl. Instrum. Meth. A} {\bf 659}, 21 (2011).
\bibitem{bes3-tau} Ablikim {\it et al.} (BESIII Collaboration), {\it Phys. Rev. D} {\bf 90}, 012001(2014).


\bibitem{pdg16} C. Patrignani {\it et al.} (Particle Data Group), {\it Chin. Phys. C} {\bf 40}, 10001 (2016).


\bibitem{gamma2} C. Bacci {\it et al.}, ($\gamma \gamma2$ Collaboration),
{\it Phys. Lett. B}{\bf 86}, 234 (1979).
\bibitem{mark1} J. L. Siegrist {\it et al.}, (Mark I Collaboration),
{\it Phys. Lett. B}{\bf 26},969 (1982).
\bibitem{pluto} L. Criegee and G. Knies, (Pluto Collaboration),
{\em Phys. Rep.} {\bf 83}, 151 (1982).

\bibitem {pluto2} Ch. Berger {\it et al.} (Pluto Collaboration), {\it Phys. Lett. B}{\bf 81},(1979).

\bibitem{alpha-err} R. Alemany, M. Davier and A. Hocker, {\it Eur. Phys. J. C} {\bf 2},123 (1998).
\bibitem{cms} S. Chatrchyan {\it et al.}, {\it Phys. Lett. B} {\bf 716}, 30 (2012).
\bibitem{atlas} G. Aad {\it et al.}, {\it Phys. Lett. B} {\bf 716}, 1 (2012).

\bibitem{qixr} X. R. Qi {\it et al.}, {\it High Energy and Nucl. Phys.} {\bf 24}, 609 (2000).
\bibitem{rvalue-bes-1} J. Z. Bai {\it et al.} (BES Collaboration), {\it Phys. Rev. Lett.} {\bf 84}, 594 (2000).
\bibitem{trigger-bes}  G. S. Huang {\it et al.}, {\it Chin. Phys. C} {\bf 25}, 889 (2001).
\bibitem{bo} Bo Andersson and H.M. Hu, hep-ph/9910285.
\bibitem{radcorr1} F.A. Berends and R. Kleiss,
{\it Nucl. Phys. B}{\bf 178}, 141 (1981).
\bibitem{radcorr2} E. A. Kuraev {\it et al.}, {\em Sov. J. Nucl. Phys.}
{ \bf 41}, 3 (1985).
\bibitem{radcorr3} G. Bonneau and F. Martin, {\it Nucl. Phys. B}{\bf 27}, 387 (1971).
\bibitem{radcorr4} C. Edwards {\it et al.}, SLAC-PUB-5160, 1990.

\bibitem{bolekpl} H. Burkhardt and B. Pietrzyk,
\Journal{\PLB}{513}{46}{2001}.
\bibitem{martin} A. Martin {\it et al.}, \Journal{\PLB}{492}{69}{2000}.
\bibitem{dave} M. Davier and A. Hoecker, \Journal{\PLB}{419}{419}{1998}.
\bibitem{kuehn} J.H. Kuehn and M. Steinhauser, \Journal{\PLB}{437}{425}{1998}.

\bibitem{rvalue-bes-3} M. Ablikim  {\it et al.} (BES Collaboration),  {\it Phys. Lett. B }{\bf 677}, 239 (2009).



\bibitem{GEANT3A} CERN Program Library Long Writeup W5013, CERN,
  Geneva, Switzerland, 1993.



\bibitem{pipiISRbabar} B. Aubert {\it et al.} (BABAR Collaboration), {\it Phys. Rev. Lett.}
{\bf 103}, 231801 (2009).
\bibitem{kloe}  F.  Ambrosino {\it et al.} (KLOE Collaboration), {\it Phys. Lett. B}
{\bf 670}, 285 (2009).
\bibitem{kloe2} F. Ambrosino et al. (KLOE Collaboration), {\it Phys. Lett. B}
{\bf 700}, 102 (2011).
\bibitem{cmd2}R. R. Akhmetshin {\it et al.} (CMD-2 Collaboration), {\it Phys.
Lett. B} {\bf 648}, 28 (2007).
\bibitem{snd}M. N. Achasov {\it et al.} (SND Collaboration), {\it J. Exp. Theor.
Phys.} {\bf 103}, 380 (2006).
\bibitem{alphaz} H. Burkhardt and B. Pietrzyk,
\Journal{\PLB}{513}{46}{2001}.

\bibitem{plb6412006145} M.Ablikim {\it et al.},  BES Collaboration, {\it Phys. Lett. B} {\bf 641},
145 (2006).
\bibitem{r-psi3770} M.Ablikim {\it et al.},  BES Collaboration, {\it Phys. Rev. Lett.} {\bf 97}, 262001 (2006).


\end{thebibliography}
\end{document}